# Magnetic properties of the RbMnPO$_4$ zeolite-ABW type material: a frustrated zigzag spin chain

G. Nénert[‡Λ], J. L. Bettis, Jr.[†], R. K. Kremer[*], H. Ben Yahia[∥], C. Ritter[‡], E. Gaudin[∥], O. Isnard[§], M.-H. Whangbo[†]

[‡] *Institut Laue-Langevin, Diffraction Group BP 156, 6 rue Jules Horowitz, F-38042 Grenoble Cedex 9, France.*

[†] *Department of Chemistry, North Carolina State University, Raleigh, North Carolina 27695-8204*

[*] *Max-Planck-Institut für Festkörperforschung, Heisenbergstrasse 1, D-70569 Stuttgart, Germany*

[∥] *Institut de Chimie de la Matière Condensée de Bordeaux (ICMCB), UPR9048-CNRS, Université Bordeaux 1, 87, Av Dr. Schweitzer, 33608 Pessac Cedex, France.*

[§] *Institut Néel, CNRS/Université Joseph Fourier, 25 rue des martyrs, BP 166, 38042 Grenoble, France.*

- [Λ]To whom correspondence should be addressed. e-mail: nenert@ill.eu

## Abstract

The crystal structure and magnetic properties of the $RbMnPO_4$ zeolite-ABW type material have been studied by temperature-dependent neutron powder diffraction, low temperature magnetometry and heat capacity measurements. $RbMnPO_4$ represents a rare example of a weak ferromagnetic polar material, containing $Mn^{2+}$ ions with $T_N = 4.7$ K. The neutron powder diffraction pattern recorded at T = 10 K shows that the compound crystallizes in the chiral and polar monoclinic space group $P2_1$ (No. 4) with the unit-cell parameters: $a = 8.94635(9)$ Å, $b = 5.43415(5)$ Å, $c = 9.10250(8)$ Å and $\beta = 90.4209(6)°$. A close inspection of the crystal structure of $RbMnPO_4$ shows that this material presents two different types of zigzag chains running along the b axis. This is a unique feature among the zeolite-ABW type materials exhibiting the $P2_1$ symmetry. At low temperature, $RbMnPO_4$ exhibits a canted antiferromagnetic structure characterized by the propagation vector $\mathbf{k_1} = \mathbf{0}$ resulting in the magnetic symmetry $P2_1'$. The magnetic moments lie mostly along the b axis with the ferromagnetic component being in the ac plane. Due to the geometrical frustration present in this system, an intermediate phase appears within the temperature range 4.7 – 5.1 K characterized by the propagation vector $\mathbf{k_2} = (k_x, 0, k_z)$ with $k_x/k_z \sim 2$. This ratio is reminiscent of the multiferroic phase of the orthorhombic $RMnO_3$ phases (R = rare earth). This suggests that $RbMnPO_4$ could present some multiferroic properties at low temperature. Our density functional calculations confirm the presence of magnetic frustration, which explains this intermediate incommensurate phase. Taking into account the strongest magnetic interactions, we are able to reproduce the magnetic structure observed experimentally at low temperature.

## Introduction

ABPO$_4$ phosphates with a large A$^{I}$ and with B$^{II}$ cations, tetrahedral coordinated, belong to a large structural family. It is characterized by a BPO$_4^{-}$ tetrahedral framework which can show four-, six- and eight-membered rings. These rings can be described by the pointing direction of the tetrahedra which are either pointing up (U) or down (D). Various topologies have been reported and some of them belong to zeolite and zeotype materials [1]. These materials have attracted a broad amount of research due to their potential applications in catalysis, ion exchange and adsorption [2]. Among the zeotype structures, the ABW framework family was first reported by Barrer and White with LiAlSiO$_4$.H$_2$O [3]. The characteristic feature of the ABW topology is the presence of sheets of six-membered rings of tetrahedra with a sequence of UUUDDD vertices within one ring. Furthermore, the ABW-zeotype offers the advantage of combining porosity with a framework structure containing potentially large amount of magnetically active species that may exhibit cooperative effects. So far, the ABPO$_4$ phosphates with the ABW-zeotype structure have been investigated mostly due to their rich polymorphism. Although many polymorphs of the ABW-zeotype are polar and can potentially contain a high content of magnetic ions, almost no investigations have been made of their magnetic properties, magnetoelectric or multiferroic properties. To the best of our knowledge, the magnetic properties of ABW-zeotype materials have been reported only for NH$_4$CoPO$_4$ (paramagnetic down to 1.7 K) [4], and KNiPO$_4$ (antiferromagnetic below T$_N$ = 25.5(5) K) [5]. In this work, we report on the complex magnetic behavior of the recently discovered RbMnPO$_4$ ABW-zeotype material [6]. By careful investigation of its crystal structure, we show that one of the two Mn chains is penta-coordinated. This is a unique feature within this family of materials. We characterize the magnetic properties of RbMnPO$_4$ by carrying out SQUID magnetometry, neutron diffraction and specific heat measurements and also by evaluating the spin exchange interactions of RbMnPO$_4$ on the basis of density functional calculations. Thanks to the combination of these methods, we show that RbMnPO$_4$ exhibits a complex magnetic behavior with an intermediate incommensurate phase stable over a very small range of temperature (4.7 < T < 5.1 K). At the lowest temperature, the system orders with a simple **k** = 0 magnetic structure (T$_N$ = 4.7 K). This behavior can be explained by competing magnetic interactions, which identifies RbMnPO$_4$ as a magnetically frustrated zigzag chain.

The article is organized as follows: after a brief description of the experimental and theoretical methods used, the results of the analysis of the crystal will be presented. Then, in the two next sections, the physical properties are described in the light of magnetic susceptibility and specific heat measurements respectively. The complex magnetic structure is investigated by neutron diffraction experiments performed at low temperature. In the last section, the obtained magnetic structure is discussed in the light of the analysis of the spin exchange interactions of RbMnPO$_4$ as derived from density functional calculation.

## Experimental Details

**Synthesis**. A powder sample of RbMnPO$_4$ was prepared by direct solid state reaction from stoichiometric mixtures of Rb$_2$CO$_3$, MnO, and (NH$_4$)H$_2$PO$_4$ powders as previously reported [6]. The mixture was fired at 500°C in argon atmosphere for 2 days and then the mixture was ground, pelletized and heated at 800$^{\circ}$C for 36 h and at 1050$^{\circ}$C for 24 h with intermediate grindings to ensure a total reaction. The resulting powder sample is very fine and light beige in color. Sample purity was checked by laboratory powder X-ray diffraction.

**Powder neutron crystallography**. Neutron diffraction data were collected on the D1A and D1B powder diffractometers at the Institut Laue Langevin using a wavelength of 1.9085 Å and 2.52 Å, respectively. The high-resolution powder diffractometer D1A is unique in being able to provide high resolution at long wavelengths, with shorter wavelength contamination eliminated by the guide tube. D1A is particularly suited to study magnetic structures and other large d-spacing studies, such as zeolites. Consequently, this was the optimal instrument for our study. The high intensity neutron powder diffractometer D1B was used to follow the detailed temperature behavior of our system taking advantage of the high counting rate available on that instrument. In this case the $\lambda/2$ contamination has been removed by using pyrolitic graphite filters whereas higher order contamination were eliminated by the neutron guide tube. The FullProf Suite program was used for nuclear and magnetic structure refinements using the Rietveld method [7].

**Magnetic measurements**. The magnetic susceptibilities were measured with a SQUID magnetometer (Quantum Design, MPMS XL) on a polycrystalline sample which was enclosed in a gelatine capsule. The magnetization of the capsule was determined in a separate run and subtracted.

**Specific heat measurements**. The specific heat was determined in a home-built adiabatic Nernst calorimeter [8] on a polycrystalline sample enclosed under 1 bar He exchange gas in a specially designed Duran glass jar. The heat capacities of the glass container and the sample platform were determined in calibration runs and subtracted.

**Density functional calculations**. We carried out spin-polarized density functional calculations by employing the projector augmented wave method encoded in the Vienna *ab initio* simulation package [9], and the generalized gradient approximation (GGA) of Perdew, Burke and Ernzerhof [10] for the exchange-correlation functionals with the plane wave cutoff energies of 400 eV, and the threshold of self-consistent-field energy convergence of $10^{-6}$ eV. To describe the electron correlation associated with the Mn 3d states, the GGA plus on-site repulsion U (GGA+U) [11] methods were applied with an effective $U^{eff} = U - J = 0$ and 2 eV on the Mn atom.

## Results and Discussion

**Crystal Structure of RbMnPO$_4$.** Among the ABW-zeotype ABPO$_4$ materials, several members crystallize in the *P2$_1$* symmetry. Namely, there are AZnPO$_4$ (A = Rb, NH$_4$) [1, 14, 15], ACoPO$_4$ (A = Rb, NH$_4$) [4], KCuPO$_4$ [12], RbCuPO$_4$ (polymorph II) [13], TlMPO$_4$ (M = Zn, Mg) and TlZnAsO$_4$ [16]. The crystal structure of RbMnPO$_4$ was successfully refined as well in the space group *P2$_1$* using the high resolution powder neutron diffraction data measured on D1A at 10 K. No impurities were detected in the neutron data. Figure 1 shows the resulting refinement. A summary of the crystallographic parameters is given in Table I. The characteristic feature of the ABW-type is the presence of sheets of six-membered rings of tetrahedra with a UUUDDD sequence of tetrahedral vertices within one ring. Subsequent layers are connected via bridging apical oxygen atoms. The structure contains channels in which Rb$^+$ cations are incorporated to achieve charge balance. A projection of the crystal structure perpendicular to the six-ring sheet is shown in Figure 2.

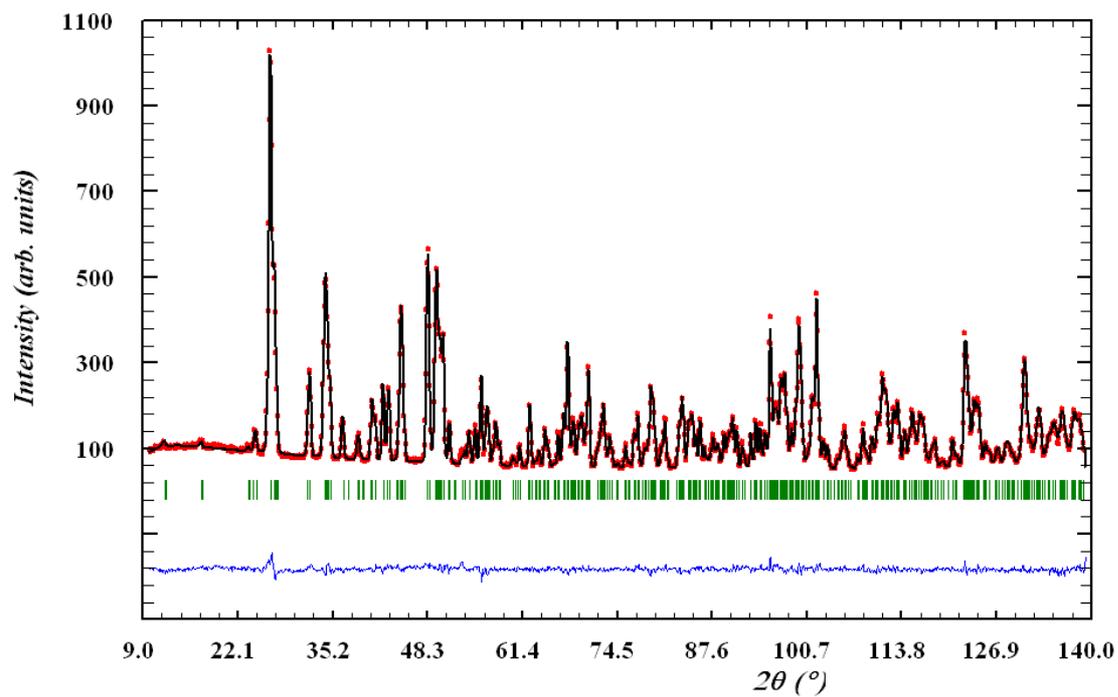

**Figure 1: Rietveld refinement of the neutron powder diffraction pattern at 10 K for RbMnPO$_4$ collected on the diffractometer D1A ($\lambda$ = 1.9085 Å). The positions of the Bragg reflections are marked by the vertical bars. The difference pattern is displayed below in blue.**

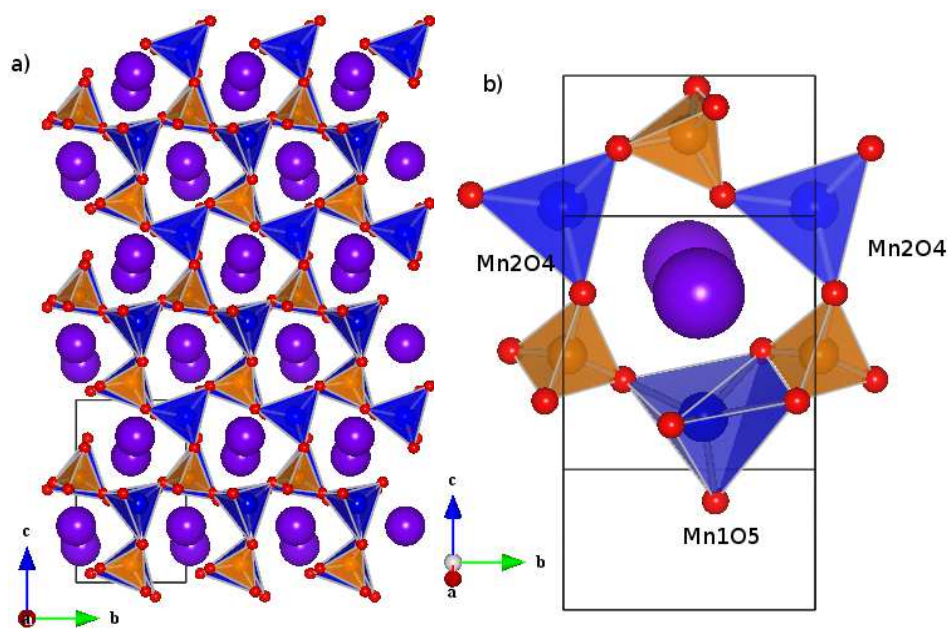

**Figure 2 : View along [100] of the crystal structure of RbMnPO$_4$ (a), and view of the six membered ring (b). Rb, Mn, P and O are shown in purple, blue, orange and red spheres, respectively. There are two manganese**

**sites Mn1 and Mn2, which are respectively penta- and tetra-coordinated by oxygen atoms. Note that two neighbouring manganese atoms are coupled through super-superexchange interactions.**

| Atoms | x/a | y/b | z/c | $U_{iso}$ |
|-------|-----|-----|-----|-----------|
| Rb1 | 0.0088(4) | 0.0379(11) | 0.1990(4) | 0.0018(5) |
| Rb2 | 0.4896(4) | 0.5000(-) | 0.6907(4) | 0.0018(5) |
| Mn1 | 0.8393(7) | 0.5553(16) | 0.4321(6) | 0.0018(5) |
| Mn2 | 0.3256(7) | 0.4946(17) | 0.0913(7) | 0.0018(5) |
| P1 | 0.7943(5) | 0.0147(12) | 0.5749(5) | 0.0018(5) |
| P2 | 0.6984(5) | 0.4962(12) | 0.0992(5) | 0.0018(5) |
| O1 | 0.5410(5) | 0.5296(13) | 0.1600(5) | 0.0053(4) |
| O2 | 0.8124(5) | 0.6038(11) | 0.2098(5) | 0.0053(4) |
| O3 | 0.7322(5) | 0.2209(11) | 0.0768(5) | 0.0053(4) |
| O4 | 0.7135(5) | 0.6331(11) | 0.9512(5) | 0.0053(4) |
| O5 | 0.9477(6) | -0.0731(13) | 0.5118(5) | 0.0053(4) |
| O6 | 0.6930(5) | 0.7921(11) | 0.5485(5) | 0.0053(4) |
| O7 | 0.7352(6) | 0.2368(12) | 0.4908(5) | 0.0053(4) |
| O8 | 0.8141(5) | 0.0767(11) | 0.7385(5) | 0.0053(4) |

Table I: Atomic coordinates of $RbMnPO_4$ determined from powder neutron diffraction at 10 K. Unit-cell parameters: a = 8.94635(9)Å, b = 5.43415(5) )Å, c = 9.10250(8)Å and $\beta$ = 90.4209(6)°. Statistics: $R_p$: 2.65%, $R_{wp}$: 3.38%, $R_{exp}$: 2.90% $\chi^2$: 1.36. The y coordinate of the Rb2 atom was fixed to define the origin of the cell.

The space group P12₁1 is a chiral space group and consequently two possible representations of the crystal structure corresponding to two different configurations can arise. By inspecting the crystal structures of RbMnPO₄ and RbCoPO₄ [4], one can notice (see Figure 3) that their structures are very similar but correspond to two different configurations. The reported configuration for RbMnPO₄ is the same as the one of the crystal structure of RbZnPO₄, TlZnPO₄ and TlZnAsO₄ [14, 16]. RbCoPO₄ exhibits an opposite configuration like the reported structure of RbCuPO₄ (polymorph II) and KCuPO₄ [4, 12, 13]. The difference between these two crystal structure types is evidenced by the difference in the orientation of the metal and the phosphorus polyhedra within the structure. The positions of phosphorus and metal ions of RbCoPO₄ are inverted with respect to those of RbMnPO₄ (see Figure 3). However, no absolute structure determination has been made on these materials with the use of resonant scattering [17] or other techniques.

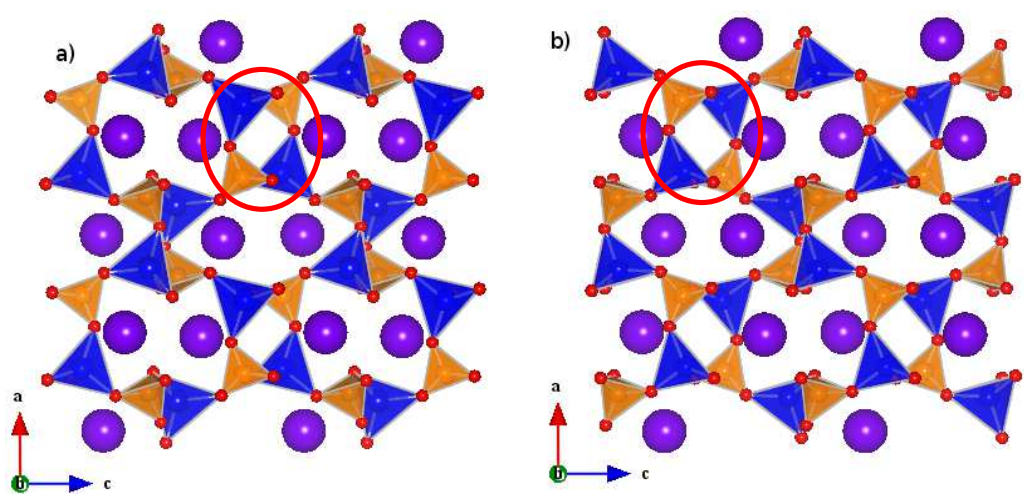

**Figure 3: Comparison of the room-temperature crystal structures of a) RbMnPO₄ and b) RbCoPO₄. The metal ion polyhedra are represented in blue, and the phosphorus polyhedra in orange. The red circles emphasize the difference between the two structures [18].**

In inorganic crystal chemistry, the structural similarity helps us to understand the conditions under which compounds of a certain composition form certain crystal structures. The structural similarity can be determined either from the general three-dimensional maps of whole structures or from some part of them. In order to further understand the particularity of $RbMnPO_4$, we compare its crystal structure to that of $RbZnPO_4$. Often a quantitative comparison of two structural models of a same phase is difficult because they are described by using different sets of atoms in the asymmetric unit and different equivalent choices of the origin or the cell orientations. One way to analyse the differences is to transform the structure of $RbMnPO_4$ to the most similar configuration of $RbZnPO_4$. The difference between the two structures is quantified by evaluating (i) the global distortion decomposed into a spontaneous strain (lattice deformation) and an atomic displacement field representing the distances between the paired atoms of the two structures and (ii) the measure of similarity as introduced by Bergerhoff *et al.* [19]. We performed this investigation using the program COMPSTRU [20] and our results are presented in table II.

| Atoms in RbZnPO$_4$ [14] | Atoms in RbMnPO$_4$ | Atomic Displacements | | | |
|---|---|---|---|---|---|
| | | $u_x$ | $u_y$ | $u_z$ | $|u|$ (in Å) |
| Rb1 | Rb1 | -0.0023 | 0.0009 | -0.0038 | 0.0397 |
| Rb2 | Rb2 | -0.0047 | -0.0003 | 0.0006 | 0.0421 |
| Zn1 | Mn1 | 0.0171 | 0.0282 | 0.0115 | 0.2373 |
| Zn2 | Mn2 | -0.0011 | 0.0225 | 0.0022 | 0.1237 |
| P1 | P1 | -0.0054 | -0.0081 | 0.0028 | 0.0698 |
| P2 | P2 | 0.0071 | 0.0290 | -0.0005 | 0.1690 |

| O1 | O1 | 0.0218 | -0.0488 | -0.0030 | 0.3283 |
| O2 | O2 | 0.0080 | 0.0194 | -0.0045 | 0.1332 |
| O3 | O3 | 0.0021 | 0.0087 | 0.0053 | 0.0692 |
| O4 | O4 | 0.0043 | 0.0305 | 0.0070 | 0.1803 |
| O5 | O5 | -0.0264 | -0.0716 | -0.0085 | 0.4581 |
| O6 | O6 | -0.0256 | -0.0104 | 0.0061 | 0.2407 |
| O7 | O7 | 0.0066 | -0.0097 | -0.0311 | 0.2903 |
| O8 | O8 | 0.0041 | 0.0097 | 0.0029 | 0.0687 |

Table II: Atom pairings and their distances for the comparison of the $RbZnPO_4$ and $RbMnPO_4$ structures. $u_x$, $u_y$ and $u_z$ are given in relative units. |u| is the absolute distance given in Å [1]

The inspection of table II shows that effectively $RbZnPO_4$ and $RbMnPO_4$ belong to an identical configurational structure type. In particular, the Rb atoms and P1 atom barely change their position going from $RbZnPO_4$ to $RbMnPO_4$. On the contrary, the M (M = Zn

---

[1] *The <u>degree of lattice distortion (S)</u> is the spontaneous strain (sum of the squared eigenvalues of the strain tensor divided by 3). For the given two structures, the **degree of lattice distortion (S)** is **0.0090**.*

*The maximum distance ($d_{max.}$) shows the maximal displacement between the atomic positions of the paired atoms. The **maximum distance ($d_{max.}$)** in this case is: **0.4581 Å***

*The <u>average distance ($d_{av.}$)</u> is defined as the average over the primitive unit cell of the distances between the atomic positions of the paired atoms. For this case the **average distance ($d_{av}$)** is calculated as **0.0801 Å**.*

*The <u>measure of similarity ($\Delta$)</u> (Bergerhoff et al., [19]) is a function of the differences in atomic positions (weighted by the multiplicities of the sites) and the ratios of the corresponding lattice parameters of the structures. The **measure of similarity ($\Delta$)** calculated for this case is **0.048**. In order to be comparable with the crystal structure of RbZnPO4, the structure of RbMnPO4 reported has been modified using a shift of -0.13020 along the b axis.*

or Co) polyhedra are significantly changed. For instance, the M1 polyhedron changes its coordination from four in RbZnPO$_4$ to five in RbMnPO$_4$. In RbZnPO$_4$, the M1 is located at a tetrahedron site made of the oxygen atoms O2, O5, O6, O7. In RbMnPO$_4$, the M1 resides in a strongly-distorted trigonal bipyramidal site. This change occurs because the M1 atom comes closer to the second symmetry related O5 atoms belonging to the next M1O$_4$ tetrahedron. Therefore this second symmetry related O5 atom belongs to both neighboring M1 polyhedra. While the M1 is displaced along the b axis towards the second symmetry equivalent O5 by about 0.15 Å (+ 0.0282 in relative units), the O5 is displaced towards M1 by about 0.38 Å (u$_y$ = -0.0716 in relative units, see table II). This coordination change characterized by the atomic displacements mostly along the b axis (including other atoms, see table II) results in the formation of Mn1O$_5$ zigzag chains running along the b direction as exemplified in Figure 4. Each Mn1O$_5$ polyhedron is turned by about 94(1) degrees relatively to the neighboring polyhedron.

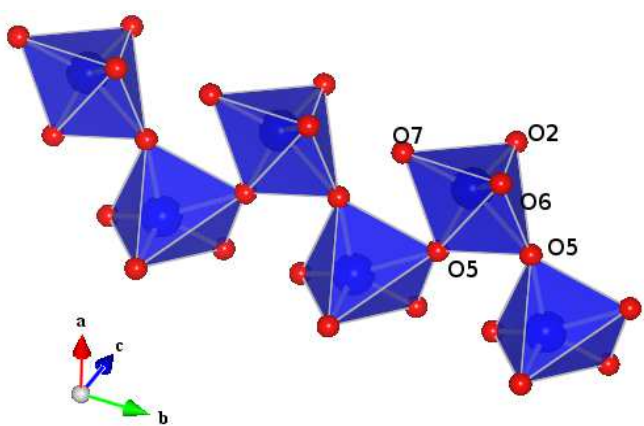

**Figure 4 : Illustration of the Mn1O$_5$ zigzag chains running along the *b* direction for the compound RbMnPO$_4$ [18]**

**Magnetic susceptibility of RbMnPO₄.** The magnetic susceptibility was measured in field cooled (FC) and zero field cooled (ZFC) modes with an applied magnetic field of 100 Oe at low temperature. An additional measurements were carried out at a higher magnetic field (H = 1000 Oe) to determine the Curie-Weiss constant and the effective magnetic moment. The obtained data are shown in Figures 5a and 5b.

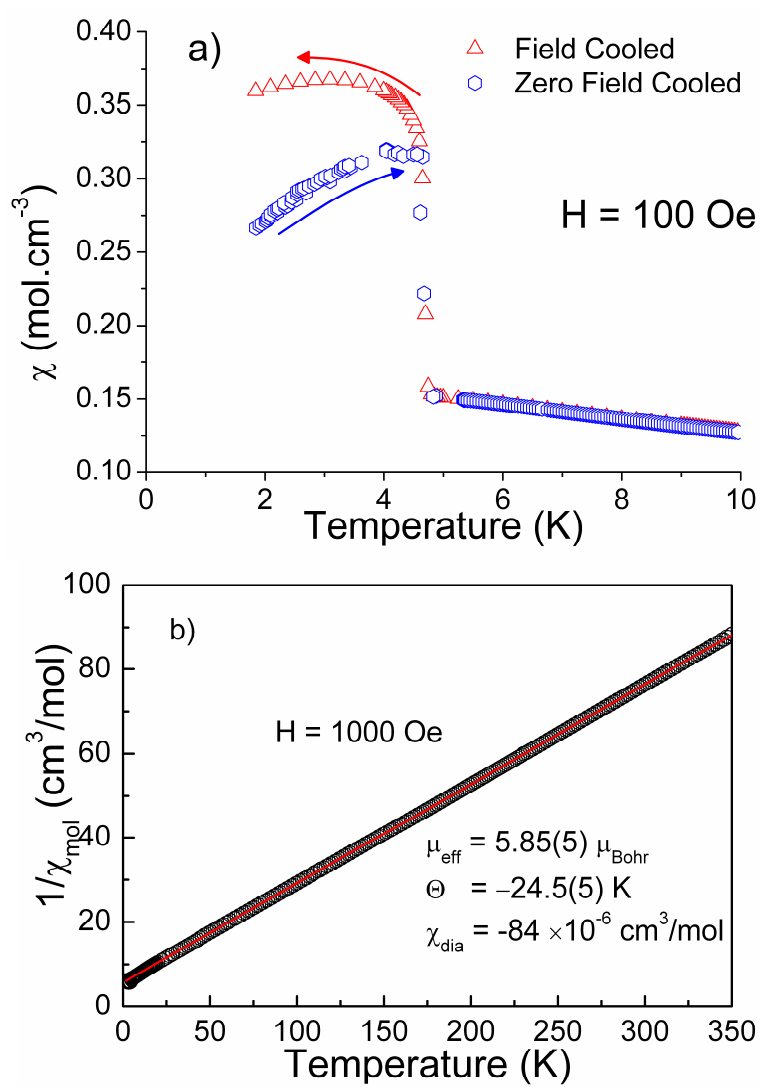

**Figure 5 :** a) Low temperature magnetic susceptibility measured at 100 Oe in zero field cooled and field cooled modes; b) Inverse magnetic susceptibility measured with H = 1000 Oe.

We fitted the magnetic susceptibility with a Curie-Weiss temperature-dependence defined by $\chi = C/(T-\theta) + \chi_0$, where the second term is the temperature-independent term. The fit was made in the range 100–350 K to find the effective moment $\mu_{eff}$ = 5.85 (5) $\mu_B$ from the Curie constant C and the Curie-Weiss temperature $\theta$ = -24.5(5) K. The negative $\theta$ indicates the presence of predominant antiferromagnetic interactions. The FC and ZFC curves exhibit a sudden increase at 4.70(5) K showing the occurrence of a single transition. These results are in qualitative agreement with previous results ($\theta$ = -27(1) K and $T_N \approx 4$ K) [6]. The ZFC curve goes through a maximum and then decreases, which suggests an antiferromagnetic order below about 4.7 K. The difference between the FC and ZFC curves indicates a canted antiferromagnetic state below 4.7 K. A magnetic system is considered to be spin frustrated when the ratio f = $-\theta/T_N$ is equal to or greater than 6. For $RbMnPO_4$, f $\approx$ 5.2, it has some magnetic frustration [21].

**Specific heat of $RbMnPO_4$.** The specific heat was measured in the temperature range of 2 to 155 K with particular attention for detailed measurement around 4.7 K. We present the specific heat as a function of temperature in a logarithmic scale in Figure 6a and in a linear scale in Figure 6b (enlargement around 4.7 K). Similar to the magnetic susceptibility, a clear anomaly is observed at T = 4.7 K. However, a further weak anomaly is clearly present just above 4.7 K, at around 5.1 K indicating the existence of an intermediate phase stable over the very narrow temperature region 4.7 < T < 5.1 K.

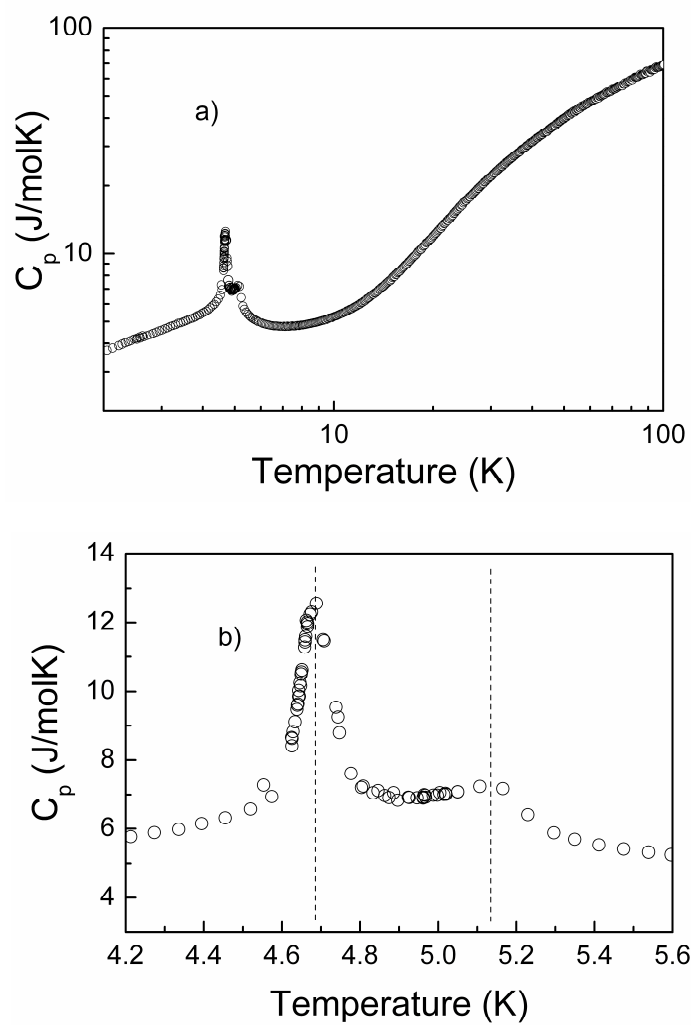

**Figure 6: Specific heat measurement of a polycrystalline sample of RbMnPO$_4$.**

**Magnetic structure from powder neutron diffraction**. The magnetic ground state of this system has been studied at low temperature by neutron diffraction using the diffractometer D1A at the Institut Laue Langevin. The magnetic Bragg reflections appearing in the powder neutron diffraction pattern can be indexed in the same unit cell as the nuclear structure using the magnetic propagation vector $\mathbf{k_1} = (0, 0, 0)$. All possible spin configurations compatible with the crystal symmetry can be derived from symmetry analysis, which has been already made in a previous study [22]. We shall recall that there are two possible magnetic structures. One is given by the magnetic symmetry *P2₁* exhibiting an antiferromagnetic coupling within the ac plane with a possible ferromagnetic component along b. The other possible magnetic structure is described by the magnetic symmetry P2₁', which has an antiferromagnetic coupling along the b axis while a ferromagnetic component is allowed within the ac plane. The better refinement of the neutron data at 2 K is obtained by using the P2₁' symmetry. The result of the Rietveld analysis with the magnetic model contribution is presented in Figure 7. We show a representation of the obtained magnetic structure in Figure 8. The refined value of the magnetic moment is 4.21(4) $\mu_B$ (the 2 non equivalent manganese sites have been constrained to have the same magnetic moment). In addition, the refinement of the magnetic structure shows the existence of a weak ferromagnetic component of about 1 $\mu_B$ per Mn site within the (ac) plane in good agreement with the SQUID data. The value of the magnetic moment is reduced compared to the nominal 5 $\mu_B$ expected for a $Mn^{2+}$ ion [23]. There are several reasons for this reduced magnetic moment. First of all, the saturation of magnetic moments is not yet reached at 2 K as evidenced by the temperature dependence investigation of the strongest magnetic Bragg peak (see below). Additionally, some magnetic frustration is present in this system as exemplified by the ratio f = $|\theta|/T_N \approx 5.2$ and confirmed by the existence of frustrated zigzag antiferromagnetic chains running along the b-direction (see below).

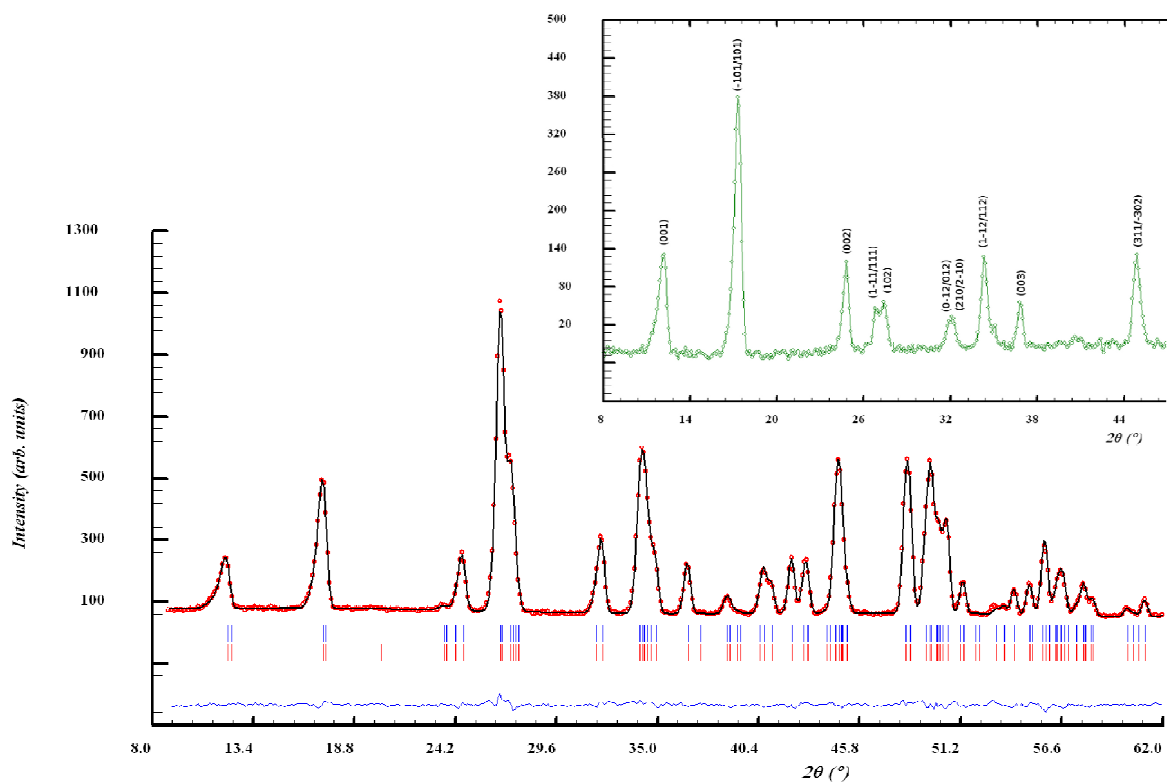

**Figure 7: Observed (red dots) versus calculated (black line) powder neutron diffraction pattern of RbMnPO$_4$ collected at 2 K on D1A ($\lambda = 1.9085$ Å). The inset shows a zoomed-in view of the difference pattern between 2 and 10 K. Statistics: R$_p$: 2.85 %, R$_{wp}$: 3.61 %, R$_{exp}$: 2.91, $\chi^2$: 1.54 & R$_{mag}$ = 2.69 %.**

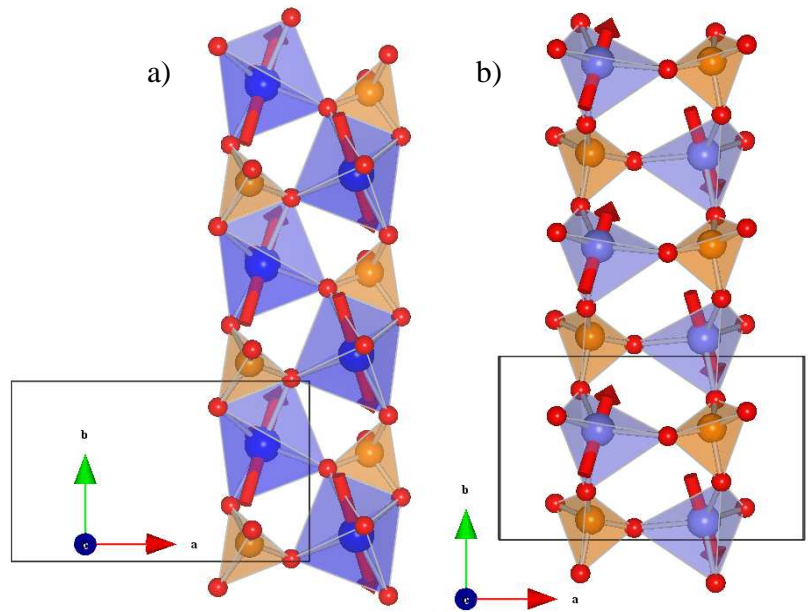

**Figure 8: Magnetic structure representation of the Néel state below $T_N$ = 4.7 K. a) The Mn1 sublattice exhibits zigzag chains with corner sharing and b) the Mn2 sublattice exhibits zigzag chains via super-superexchanges only [18].**

We have further investigated the temperature dependence of the magnetic structure by performing short scans while warming up through the magnetic transitions. These data were collected using the low resolution, high intensity powder diffractometer D1B. While we were interested in the temperature development of the Néel state below $T_N$ = 4.7 K, we aimed also at investigating the nature of the intermediate phase stable between 4.7 and 5.1 K. We show in Figure 9 a portion of the powder neutron pattern as function of temperature, which clearly shows weak extra scattering intensity at low angles between 4.7 and 5.1 K corresponding to an incommensurate propagation vector $\mathbf{k_2}$ = ($k_x$ 0 $k_z$) with $k_x$ = 0.1515(4) and $k_z$ = 0.0648(5) at 4.7 K. This new propagation vector is temperature-dependent. However due to the weakness of these reflections and the broad uneven background, we were unable to describe the data with an appropriate model.

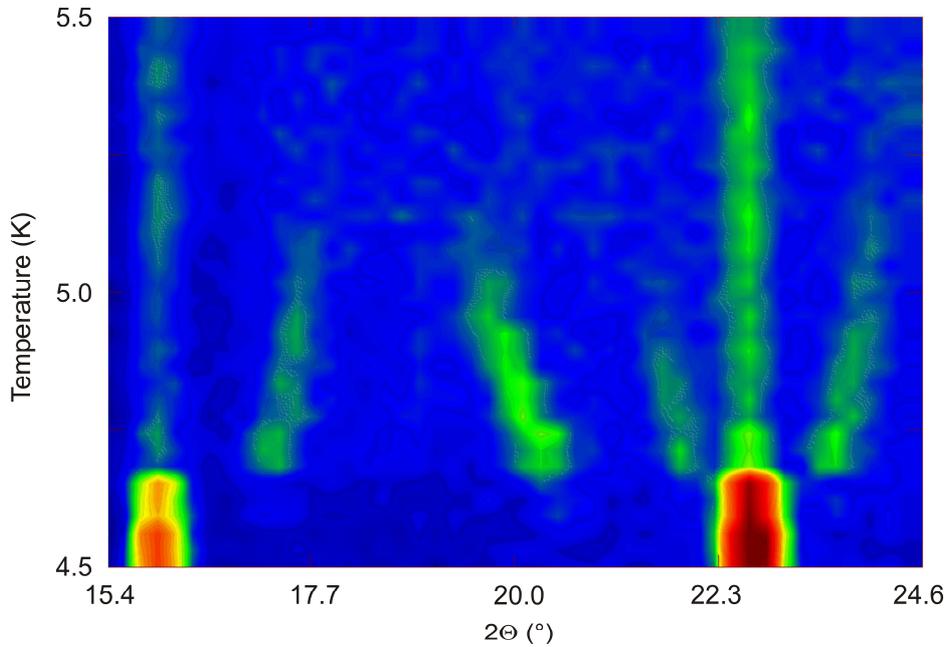

**Figure 9: Portion of the powder neutron pattern of RbMnPO$_4$ as function of temperature in the range 4.5 – 5.5 K as measured on the diffractometer D1B. Satellite reflexions origin are observed before the appearance of magnetic contribution on top of the nuclear scattering below about 4.7 K.**

The temperature dependence of the diffraction pattern presented in Figure 9 highlights the temperature development of the propagation vector $\mathbf{k_2}$. In Figure 10, we plotted the evolution of the values of the coordinates of $\mathbf{k_2}$. The values of $k_x$ and $k_z$ evolve with temperature to finally lock into the commensurate values $k_x = k_z = 0$ at about 4.7 K. We notice that we keep the ratio $k_x/k_z \approx 2$ over the whole temperature range. A first interpretation of the reflections appearing within this narrow temperature range is that they are magnetic in origin. However, the constant 2 ratio of the vector components reminds us of the multiferroic orthorhombic RMnO$_3$ phases. There, accompanying the magnetic ordering, a lattice modulation with $\delta_l = 2\delta_m$ ($\delta_l$ lattice modulation and $\delta_m$ magnetic modulation) is also observed [24]. Consequently an alternative interpretation of the data would be that $k_x$ is of structural origin while $k_z$ would be of magnetic origin. In order to disentangle these 2 different scenarios, low temperature synchrotron radiation and/or polarized neutron scattering experiment on single crystals are necessary. If $k_x$ is of structural origin, this implies that we have a magnetically induced ferroelectric state between 4.7 and 5.1 K with a re-entrant polar state below 4.7 K due to the *P2₁* symmetry. Finally, we note that the two transitions at 4.7 and 5.1 K have a strong first order character.

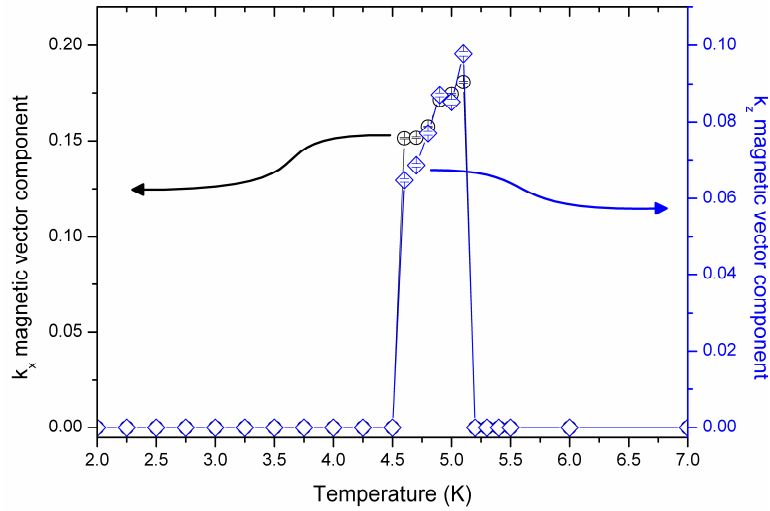

**Figure 10: Temperature dependence of the vector components $k_2$ as a function of temperature.**

In addition to this intermediate phase, we were able to follow the temperature dependence of the magnetic structure of the **k** = 0 magnetic phase. Figure 11 presents the integrated intensity of the (-101/101) peak as function of temperature. A fit of the data to a power law $I = I_0(T_N-T)^\beta$ (illustrated by the solid red line in Figure 11) reveals a critical exponent $\beta = 0.38(1)$, which is consistent with typical values (0.3645(25) [25]) found for three-dimensional systems. This suggests that below $T_N = 4.7$ K RbMnPO₄ behaves like a 3D system despite the presence of zigzag chains. Consequently, one can deduce that the inter-chain exchange interactions are not negligible in such system.

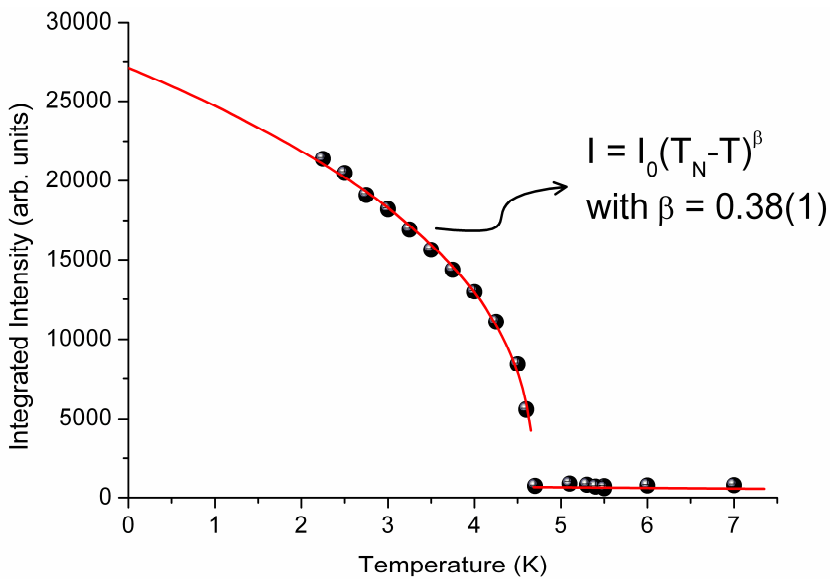

**Figure 11: Integrated Intensity of the (-101/101) magnetic peak as function of temperature. The line corresponds to a fit to the power law I = I$_0$(T$_N$-T)$^\beta$ below T$_{N2}$~4.7 K and constant above.**

## Analysis of the Spin exchange interactions of RbMnPO$_4$ below T$_N$ = 4.7 K

RbMnPO$_4$ contains high spin Mn$^{2+}$ (d$^5$, S = 5/2) and undergoes an antiferromagnetic ordering below T$_N$ = 4.7 K. To examine the nature of the spin exchange interactions in RbMnPO$_4$, we evaluate the 10 spin exchange parameters, J$_1$ – J$_{10}$, on the basis of DFT calculations. The structural parameters associated with these spin exchange paths are summarized in **Table III**, and their 3D arrangement is presented in **Figure 12**. We determined the relative energies of the 11 ordered spin states shown in **Figure S1** of the supporting information (SI) constructed by using a (2a, 2b, 2c) supercell. The relative energies of these 11 ordered spin states determined for the room-temperature structure of RbMnPO$_4$ by GGA+U calculations (with U = 0 and 2 eV) are summarized in **Table S1** of the SI. In terms of the Heisenberg spin Hamiltonian, $\hat{H} = -\sum_{i<j} J_{ij} \hat{S}_i \cdot \hat{S}_j$, where $J_{ij} = J_1 - J_{10}$, we determine the total spin-exchange energies per (2a, 2b, 2c) supercell (i.e., per 32 formula units) of the 11 ordered spin states by applying the energy expression obtained for spin dimers with N unpaired spins per spin site (in the present case N = 5) [26-28]. The results are summarized in the SI. The relative energies of the 11 ordered spin states determined by GGA+U calculations are mapped onto the corresponding relative energies determined from the above spin-exchange energies to find the values of $J_1 - J_{10}$ summarized in **Table IV**. It is worth noting that the interchain interactions $J_2$, $J_3$ and $J_5$ are not negligible in comparison with the intra chain interactions. They are of the same order of magnitude. These calculations confirm the 3D character at the low temperature magnetic structure.

**Table III**. Geometrical parameters associated with the spin exchange paths $J_1 - J_{10}$ in the structure of $RbMnPO_4$, where the $MnO_n$ (n = 4 or 5) polyhedra are indicated by the numbers n, and the Mn…Mn and the shortest O…O distances are given in Å.

|          | polyhedra | Mn…Mn/ O…O |
|----------|-----------|--------------------|
| $J_1$    | (5-4)     | 5.560/2.477        |
| $J_2$    | (5-4)     | 5.209/2.731        |
| $J_3$    | (5-4)     | 5.471/2.831        |
| $J_4$    | (5-5)     | 4.097/  -          |
| $J_5$    | (5-4)     | 5.638/2.486        |
| $J_6$    | (5-4)     | 5.880/2.539        |
| $J_7$    | (4-4)     | 4.483/2.226 & 2.554 |
| $J_8$    | (5-4)     | 5.352/2.629        |
| $J_9$    | (5-5)     | 5.451/2.633        |
| $J_{10}$ | (4-4)     | 5.451/2.658        |

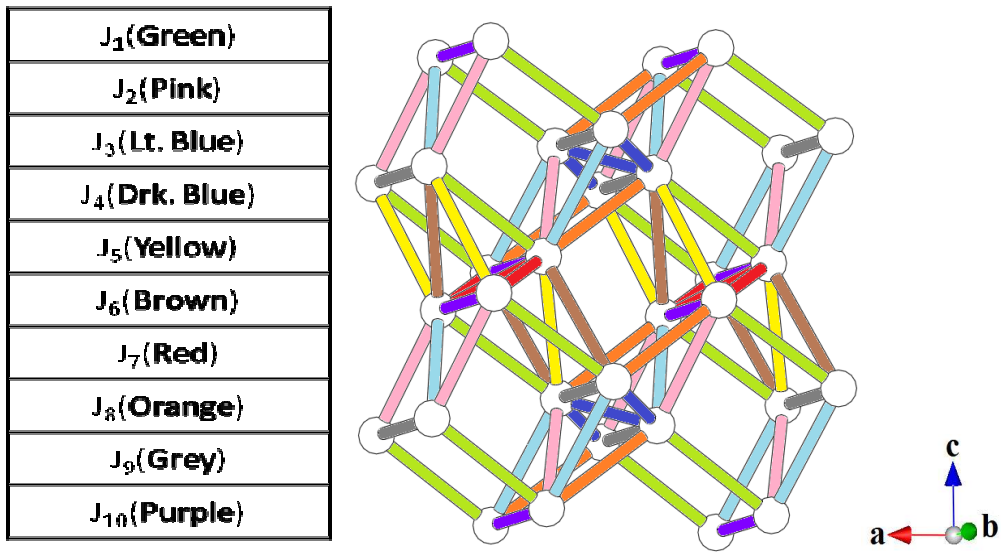

| $J_1$(Green) |
| $J_2$(Pink) |
| $J_3$(Lt. Blue) |
| $J_4$(Drk. Blue) |
| $J_5$(Yellow) |
| $J_6$(Brown) |
| $J_7$(Red) |
| $J_8$(Orange) |
| $J_9$(Grey) |
| $J_{10}$(Purple) |

**Figure 12**. 3D arrangement of the $Mn^{2+}$ ions (empty spheres) and the spin exchange paths (colored cylinders) in $RbMnPO_4$.

**Table IV**. Values of $J_1 - J_{10}$ (in units of K) extracted for the structure of $RbMnPO_4$ from GGA+U calculations.

|          | $U = 0$ eV | $U = 2$ eV |
|----------|------------|------------|
| $J_1$    | -0.74      | -0.41      |
| $J_2$    | -6.69      | -4.06      |
| $J_3$    | -1.28      | -0.74      |
| $J_4$    | -9.64      | -5.17      |
| $J_5$    | -2.13      | -1.09      |
| $J_6$    | -1.90      | -1.01      |
| $J_7$    | -6.77      | -3.82      |
| $J_8$    | -1.70      | -1.05      |
| $J_9$    | -3.46      | -1.89      |
| $J_{10}$ | -2.61      | -1.55      |

Table IV shows that the spin exchange parameters calculated with U = 0 are greater in magnitude than those calculated with U = 2 eV. To determine which set of the $J_1 - J_{10}$ values is more appropriate for our discussion, we calculate the Curie-Weiss temperature $\theta$ by using the calculated $J_1 - J_{10}$ values. Since there are two different Mn sites (i.e., Mn1 and Mn2) in equal number, we calculate the Curie-Weiss temperature as follows:

$$\theta = [\theta(Mn1) + \theta(Mn2)] / 2 \tag{1}$$

where

$$\theta(Mn2) = [S(S+1)/3](J_1 + J_2 + J_3 + J_5 + J_6 + 2J_7 + J_8 + 2J_{10})$$

$$\theta(Mn1) = [S(S+1)/3](J_1 + J_2 + J_3 + 2J_4 + J_5 + J_6 + J_8 + 2J_9)$$

$$\tag{2}$$

The Curie-Weiss temperature derived from the $J_1 - J_{10}$ values is $\theta$ = -107.7 K from the GGA+U calculations with U = 0, and $\theta$ = -60.6 K from those with U = 2 eV. Given the experimental $\theta$ = -24.5(5) K, the $J_1 - J_{10}$ values obtained from the use of U = 2 eV are more reasonable.

We note from **Table IV** that the spin exchanges $J_2$, $J_4$, $J_7$, and $J_9$ are the four strongest AFM spin exchanges. As illustrated in **Figure 13**, the spin exchange interactions show that the two strongest spin exchanges $J_4$ and $J_7$ each form zigzag AFM chains along the b-direction. The $J_4$-zigzag AFM chain is spin-frustrated by $J_9$. The $J_7$-zigzag AFM chain is spin-frustrated by $J_{10}$. On the basis of the spin exchange interactions, we now consider a possible 3D AFM structure expected for RbMnPO$_4$ below $T_N$ = 4.7 K. As shown in **Figure 14**, the $J_4$- and $J_7$-zigzag AFM chains along the b-direction can be antiferromagnetically ordered via the interchain AFM interactions $J_2$, $J_3$ and $J_5$ to form a 3D AFM lattice. The unit cell of this 3D magnetic structure is identical with the chemical unit cell of RbMnPO$_4$. This prediction is consistent with the experimental observation.

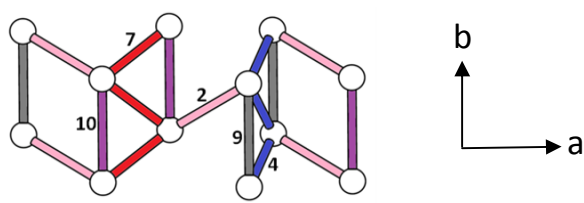

**Figure 13.** Five strongest spin exchange interactions present in the structure of RbMnPO$_4$, where the numbers n (e.g., 2, 4, 7, 9, 10) represent the spin exchange J$_n$.

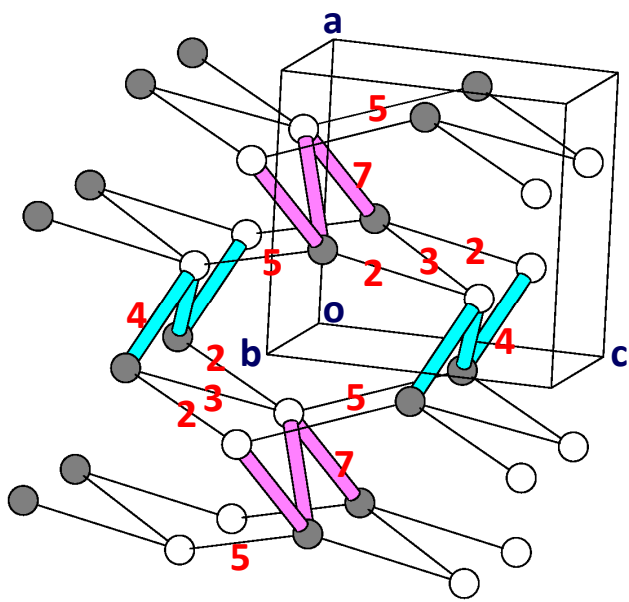

**Figure 14.** 3D AFM order of RbMnPO$_4$ expected below T$_N$ = 4.7 K. The filled and empty circles represent the up-spin and down-spin Mn$^{2+}$ sites, respectively. The strong spin exchange interactions J$_4$ (cyan magnetic bonds) form 1D AFM chains along the b-direction, and so do the strong spin exchange interactions $J_7$ (magenta magnetic bond). These chains interact antiferromagnetically via the interactions $J_2$, $J_3$ and $J_5$ to form a 3D AFM lattice. The magnetic unit cell of this 3D magnetic structure is the same as the chemical unit cell of $RbMnPO_4$. The numbers n (e.g., 2, 3, 4, 5, 7) represent the spin exchange $J_n$.

## Conclusion

The crystal structure and magnetic properties of the $RbMnPO_4$ zeolite-ABW type material have been studied by temperature-dependent neutron powder diffraction, superconducting quantum interference device (SQUID) magnetometry and heat capacity measurements. $RbMnPO_4$ represents a rare example of a weak ferromagnetic polar material, containing $Mn^{2+}$ ions with $T_N$ = 4.7 K. We have shown that $RbMnPO_4$ presents the unique feature of having two different types of zigzag chains exhibiting magnetic frustration. Due to the geometrical frustration present in this system, an intermediate phase appears within the temperature range 4.7 – 5.1 K characterized by the propagation vector $k_2 = (k_x, 0, k_z)$ with $k_x/k_z \sim 2$, reminiscent of the multiferroic phase of the orthorhombic $RMnO_3$ phases (R = rare earth). Our density functional calculations confirm the presence of magnetic frustration, which explains this intermediate incommensurate phase. Taking into account the strongest magnetic interactions, we are able to reproduce the magnetic structure observed experimentally at low temperature. Both experimental results and theoretical calculations agree to indicate that the interchain exchange coupling constants are non negligible and play a significant role in the 3D character of the magnetic ground state of $RbMnPO_4$.

**ASSOCIATED CONTENT**
**Supporting Information**

Derivation and relative energies (in meV/FU) of the 11 ordered spin states and their associated representations of $RbMnPO_4$ obtained from GGA+U calculations. This material is available free of charge via the Internet at http://pubs.acs.org."

**Acknowledgments**

At NCSU this research was supported by the computing resources at the NERSC and the HPC Centers. The authors thank the Institut Laue Langevin for the allocation of beamtime and technical support.

# References

[1] Breck D. W., *Zeolite Molecular Sieves*; Wiley & Sons: New York, 1974; Kahlenberg V., Fischer R. X., Baur W. H.; Z. Kristallogr. 216, 489 (2001); Bu X., Feng P., Gier T. E., Stucky G. D., Zeolites 19, 200-208 (1997), Gier T.E., and Stucky, G.D. Nature, 349, 508-510 (1991); Weller M. T., J. Chem. Soc., Dalton Trans., (2000), 4227-4240

[2] Dong J., Wang X., Xu H., Zhao Q., Li J.; International Journal of Hydrogen Energy 32 (2007), 4998-5004;

[3] Barrer R. M. and White E. A. D., J. Chem. Soc., 1267 (1951)

[4] Feng P., Bu X., Tolbert S. H., Stucky G. D., J. Am. Chem. Soc., 119 2497-2504 (1997)

[5] Fischer P., Luján M., Kubel F., Schmid H., Ferroelectrics vol. 162, 37-44 (1994)

[6] Ben Yahia H., Gaudin E., Darriet J., Journal of Alloys and Compounds, Vol. 442, Pages 74–76 (2007)

[7] Rodriguez-Carvajal J., Physica B **192**, 55 (1993).

[8] Schnelle W., Gmelin E., Thermochim. Acta, vol. 391, 41 (2002)

[9] (14) (a) Kresse, G.; Hanfner, J. *Phys. Rev. B* **1993**, *47*, *558*. (b) Kresse, G.; Furthmúller, J. *Comput. Mater. Sci.* **1996**, *6,* 15. (c) Kresse, G.; Furthmúller, J. *Phys. Rev. B.* **1996**, *54,* 11169.

[10] Perdew, J. P.; Burke, S.; Ernzerhof, M. *Phys. Rev. Lett.* **1996**, *77,* 3865.

[11] Dudarev, S. L.; Botton, G. A.; Savrasov, S. Y.; Humphreys, C. J.; Sutton, A. P. *Phys. Rev. B* **1998**, *57*, 1505.

[12] Shoemaker G. L., Kostiner E., Anderson J. B., Zeitschrift für Kristallographie 152, 317-332 (1980)

[13] Henry P. F., Kimber S. A. J., Argyriou D. N., Acta Cryst. B66, 412-421 (2010)

[14] Elammari L., Elouadi B., J. Chim. Phys. 88, 1969-1974 (1991)

[15] Averbuch-Pouchot M.T., Durif A., Materials Research Bulletin 3, 719-722 (1968).

[16] Andratschke M., Range K.- J., Weigl C., Schiessl U., Rau F., Zeitschrift für Naturforschung B, 49, 1282-1288 (1994)

[17] Bijvoet J. M., Peerdeman A. F. & van Bommel A. J., *Nature* **168**, p 271 (1951)

[18] Momma K. and Izumi F., J. Appl. Crystallogr. **41**, 653 (2008).

[19] Bergerhoff G., Berndt M., Brandenburg K., Degen T., Acta Cryst. B 55, 147-156 (1999)

[20] Tasci E. S., de la Flor G., Orobengoa D., Capillas C., Perez-Mato J.M., Aroyo M. I., EPJ Web of Conferences **22** 00009 (2012).

[21] Ramirez A. P., Annu. Rev. Mater. Sci. **24**, 453 (1994)

[22] Nénert G., Koo H.-J., Colin C. V., Bauer E. M., Bellitto C., Ritter C., Righini G., Whangbo M.-H..; *Inorg. Chem. (2013), 52, 753–7606*

[23] Jacobson A. J., Tofield B. C., Fender B. E. F., J. Phys. C: Solid State Phys. 6, 1615 (1973)

[24] Kimura T., Goto T., Shintani H., Ishizaka K., Arima T., Tokura Y., *Nature* **426**, 55-58 (2003)

[25] Le Guillou J. C., Zinn-Justin J., Phys. Rev. B 21, 3976 (1980)

[26] Dai D. and Whangbo M.-H., *J. Chem. Phys.* 114 (7), 2887 (2001); *J. Chem. Phys*. **118**, 29 (2003).

[27] Whangbo M.-H., Koo H.-J. and Dai D., *J. Solid State Chem*. **176**, 417 (2003).

[28] Xiang H. J., Lee C., Koo H.-J., Gong X. G. and Whangbo M.-H., *Dalton Trans.* **42**, 823 (2013).

Table of Content

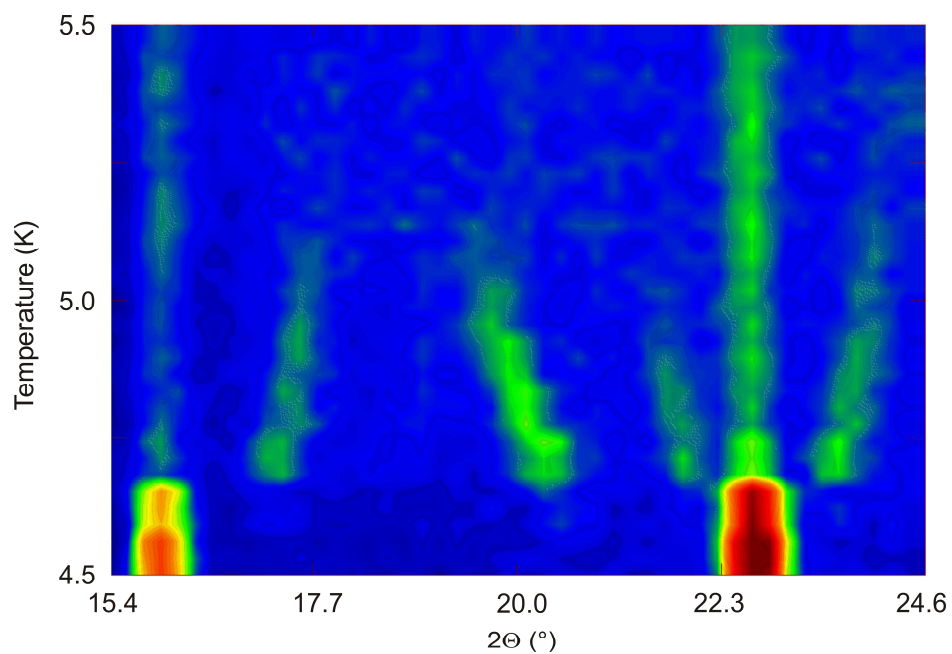

RbMnPO$_4$ is a rare example of a polar weak ferromagnet exhibiting a Néel state below T$_N$ = 4.7 K. In addition, this material shows magnetic frustration resulting in the emergence of an intermediate phase reminiscent of the multiferroic phase reported in RMnO$_3$. Theoretical analysis based on first-principles DFT calculations details the various magnetic exchange interactions explaining the observed complex behavior and ground state of the system.